\documentclass[prc,onecolumn,tightenlines,12pt]{revtex4}
\usepackage{graphicx}
\usepackage{dcolumn}
\usepackage{amsmath}
\newcommand{\be}{\begin{equation}}
\newcommand{\ee}{\end{equation}}
\newcommand{\bea}{\begin{eqnarray}}
\newcommand{\eea}{\end{eqnarray}}

\begin{document}

\author{J. X. de Carvalho$^{1}$, 
M. S. Hussein$^{1,2}$, and Weibin Li$^{1}$} 

\affiliation{$^{1}$Max-Planck-Institut f\"ur Physik komplexer Systeme\\
N\"othnitzer Stra$\beta$e 38, D-01187 Dresden, Germany \\
$^{2}$Instituto de F\'{i}sica, Universidade de S\~{a}o Paulo\\
C.P. 66318, 05315-970 S\~{a}o Paulo, S.P., Brazil}
\title{Quantum reflection: The invisible quantum barrier}

\begin{abstract}
We construct the invisible quantum barrier which represents the phenomenon of
quantum reflection using the available data. We use the Abel equation to invert
the data. The resulting invisible quantum barrier is double-valued in both axes.
We study this invisible barrier in the case of atom and Bose-Einstein Condensate reflection 
from a solid silicon surface. A time-dependent, one-spatial dimension Gross-Pitaevskii equation
is solved for the BEC case. We found that the BEC behaves very similarly to the single atom
except for size effects, which manifest themselves in a maximum in the reflectivity at small
distances from the wall. The effect of the atom-atom interaction on the BEC reflection and correspondingly 
on the invisible barrier is found to be appreciable at low velocities and comparable to the finite size effect.
The trapping of ultracold atom or BEC between two walls is discussed.
\end{abstract}


\maketitle

\section{Introduction}
Quantum reflection is an intriguing phenomenon which has been under experimental scrutiny
in the last few years. Several papers have been written on the subject. Back in 1983, V. U. Nayak,
D. O. Edwards and N. Masuhara \cite{Na83} measured the scattering of $^{4}$He atoms grazing the liquid-$^{4}$He
surface and found that the reflectivity approaches unity as the velocity tends to zero. Berkhout et al. \cite{Be89}
measured the quantum reflection of H atoms from a concave spherical mirror. Yu et al.\cite{Yu93} revisited the
experiment of \cite{Na83} and measured the sticking probability. More recently, \cite{Sh01} looked at the phenomenon 
of quantum reflection of very slow metastable neon atoms from a solid silicon surface and a BK7 glass surface. 
More data were collected in the following years. 
Sodium Bose-Einstein condensate (BEC) containing $3 \times 10^5$ atoms reflected at normal incidence from a silicon surface
at incident velocities of 1 - 8 mm/s were reported by \cite{Pas04}. Extension of this study to quantum reflection of a $^{23}$Na BEC containing $10^6$ atoms at velocities below 2.5 mm/s on a dilute silicon surface demonstrated the effect of the atom-atom interaction
and the extended density profile as the reflectivity saturated at 0.6 rather than 1.0 \cite{Pas06}. The observed 
reflectivity is explained as quantum reflection caused by the attractive 
Casimir-Polder (CP) atom-wall potential \cite{CP48} The significant reduction in the reflectivity of BEC at very low velocities allegedly \cite{Sco05} arises from the modulation in the BEC density profile which ensues owing to the formation of standing waves resulting from the superposition of incident and reflected matter waves. As a result, dynamical excitations, such as solitons and vortex rings, are created in the BEC which fragment it and disperses its atoms. Such work on quantum reflection of BEC's is important for the understanding of the Bose-Einstein condensate stability near surfaces \cite{Lin04}. Several recent theoretical works on the general foundation of quantum reflection have been published
\cite{ MH96, Hen96, Seg97, Fri02}.

\bigskip

The observed reflectivity is explained as quantum reflection caused by the \emph{attractive}
Casimir-Polder potential \cite{CP48}. The reflection is accordingly classically forbidden. The Wentzel-Kramers-Brillouin (WKB) approximation would not work in such a case since the variation of the local de Broglie wavelength is comparable to itself, regardless to the sign of the interaction. This, however,
does not preclude the use of adequate perturbation methods to calculate very approximately the reflectivity \cite{MH96}. On the other hand, reflection implies tunneling and one can thus invert the data on reflectivity, $R(E)$ (or tunneling, $T(E)$ = 1 - $R(E)$) using WKB type formulae \cite{CG78}, to obtain the width of what might be called the invisible quantum barrier. Such a procedure would avoid the use of ill-defined concepts such as the quantum potential \cite{MH96}. The purpose of this paper is to use the inversion procedure to obtain the invisible barrier responsible for quantum reflection. We apply this to the available data on quantum reflection of atoms \cite{Sh01} and of BEC \cite{Pas04, Pas06}. We also consider the resulting invisible barrier in the context of the trapping of BEC between two walls. \\

The paper is organized as follows. In section II, the phenomenon of Quantum Reflection is reviewed and the Casimir-Polder potential is discussed. In Section III, the Invisible Barrier is derived using the Abel equation for existing data on atom-wall and BEC-wall reflection. In Section IV, the Invisible Barrier is obtained for a BEC using a one dimensional Gross-Pitaevskii equation. The free space trapping of BEC between two walls is also considered. The trapping occurs within the confines of the two invisible barriers which sit at an appreciable distance far from the walls that create them. Finally, in Section V, several concluding remarks are made. 

\bigskip

\section{Quantum Reflection}

The phenomenon of quantum reflection occurs whenever the local de Broglie wave length of the moving particle is comparable
to the distance over which the potential varies rapidly, regardless to whether the potential is repulsive or attractive. For an attractive potential, such as the one between an atom and a wall ( the Casimir-Polder interaction), the quantum reflection is a classically forbidden process, just like tunneling through a barrier. In this section we give a short account of Quantum Reflection, in connection with the CP interaction. \\

The CP, atom-wall, interaction, which has the same physical vacuum fluctuation origin as the famous Casimir force between parallel walls separated by a distance $L$, , $V_{C}(r) = -\frac{\hbar c\pi^2}{240}\frac{1}{L^4}$,  is derived in \cite{CP48} and can be written as an integral ( see, e.g. \cite{Mar97, Kha97}),

\begin{equation}
V_{CP}(r) = -\frac{1}{4\pi\alpha_{fs}r^4}\int_0^\infty~dx~\alpha(\frac{ix}{\alpha_{fs}r})\exp(-2x)\big[2x^2+2x+1\big]
\end{equation}

where $\alpha(i\omega)$ is the dynamic electric dipole polarizability of the atom evaluated at the imaginary frequency i$\omega$ and $\alpha_{fs}$ is the fine-structure constant. The quantity $\alpha(i\omega)$ can be evaluated using sum rules and it can be written as,

\begin{equation}
\alpha(i\omega) =  S_{n}\frac{f_{n}}{(E_{n} -E_{0})^2 + \omega^2}
\end{equation}

where the sum $S_{n}$ runs over the discrete dipole states and the continuum scattering states. In the above, the dipole oscillator strength $f_{n}$ is given by

\begin{equation}
f_{n} = \frac{2m_{e}}{3\hbar^2} (E_{n} -E_{0}) \Bigg|\Bigg<0\Bigg|\sum_0^N \mathbf{r_{i}}\Bigg|n\Bigg>\Bigg|^2
\end{equation}

In the above equation N represents the number of electrons. Further, the quantity $S_{n}f_{n}$ is just the dipole sum rule $\frac{m_{e}}{\hbar^2}N$, \cite{Kha97}.

The asymptotic form of the CP potential is invariably written as,

\begin{equation}
V_{CP}(r) = \frac{1}{r^3}\frac{-C_{4}}{(r + \frac{3\lambda}{2\pi^2})}
\end{equation}

where $C_{4}$ has the value,e.g., $9.1\times 10^{-56}$ $J.m^{4}$ for sodium atoms incident on a silicon surface and $\lambda = 590$ nm, \cite{Pas04,Pas06}

The result of the Quantum Reflection measurements cited above show a strikingly robust quantum reflection at very low velocities. The presence of such quantum reflection raises the question about the transmission to the solid wall, where the atoms are adsorbed.
The transmission, or tunneling implies the presence of an invisible quantum barrier "sitting" at a distance close to the
wall. Can one determine this barrier from the available data
on quantum reflection? How does a BEC behave compared to a single atom? The answer to these queries is the thrust of the present paper.

Knowing the reflection coefficient, the transmission one is simply
the compliment to unity, viz $T(E) = 1 - R(E)$.

In the following we use the generalized WKB form for the tunneling probability, $T(E)$,\cite{Kem35}

\begin{equation}
T(E) = \frac{1}{1 + \exp\big(2\sqrt\frac{2m}{\hbar^{2}}\int_{r_{1}}^{r_{2}}~dr\sqrt{V(r) -E}~\big)}
\end{equation}

Above, $r_{1}$ and $r_{2}$ are the inner and outer turning points
which are the roots of the equation $V(r) = E$. Further, the above form of the tunneling probability is exactly 1/2 at the top
of the barrier as the exact solution of the Schr\"odinger equation so requires.

Knowing $T(E)$ from the data, one can invert it to obtain the barrier. This procedure is intimately related
to the solution of Abel's theorem and the corresponding classical mechanics problem \cite{LL69}. The equation that does this is given by

\begin{equation}
r_{2} - r_{1} = \frac{-1}{\pi}\frac{\hbar}{\sqrt{2m}}\int_V^B dE~\frac{R^{\prime}(E)}{R(E)T(E)}\frac{1}{\sqrt{E-V(r)}}
\end{equation}
where $B$ is the height of the barrier, generally defined through $T(B)= R(B)=\frac{1}{2}$, which results from $\int_{r_{1}}^{r_{2}}~dr\sqrt{V(r)- B}= 0$. This last relation defines theoretically the height of the barrier, while the former supplies an "experimental" definition of this height.

The above equation, which we shall refer to as Abel formula, has been used in a variety of forms to investigate inversion in the WKB type description of bound systems by
\cite{CG78}. It has also been used to extract the barrier responsible for tunneling into the classically forbidden region
in the sub-barrier fusion of nuclei \cite{BKN83}. The resulting potential which results from the inversion of the data is multivalued function. Of course, by its nature, Eq. (6) dictates that the thickness function, $r_{2} - r_{1}$, is determined
by $R(E)$, or equivalently, $T(E)$ for $ V < E < B$, namely in the energy region where $R(E) \geq 1/2$.

\bigskip

\section{Invisible Quantum Barrier}
Cole and Good \cite{CG78}have suggested using the Abel formula to invert the semiclassical reflectivity and tunneling probability to obtain the thickness of the barrier in cases of reflection in the classically allowed region. Here we have used the Abel formula to obtain what we may call the invisible quantum barrier (INV) responsible for the 
quantum reflection (QR), reflection in the classically forbidden region, of single sodium atoms \cite{Sh01} from a silicon surface, and for the QR of a Bose-Einstein 
condensate composed of a 3 x $10^{5}$ sodium atoms in a trap \cite{Pas04} and of a million such atoms \cite{Pas06}. Although the interaction between BEC and a surface is less well understood than that of a single atom with a surface, we take the practical view of using the same type of the CP interaction as was done in \cite{Pas04, Pas06, Sco05}. When using Eq. (6), we have fitted the data points with an analytical function of the velocity and used this function in the inversion. Of 
course the inversion alluded to above would only supply the barrier value as a function of its thickness, $V_{inv} ( x_{1}-x_{2})$. Of course this procedure is applicable at energies where the reflectivity is larger than $\frac{1}{2}$.\\ 

For energies greater than the height of the barrier, $R(E) < \frac{1}{2}$, one can use the parabolic barrier approximation valid near the position, $R_{B}$, of the height, $B$,  of the barrier, $V_{inv}(r) \approx B - \frac{1}{2}|\frac{d^{2}V_{inv}}{dr^2}|_{r = R_{B}} (r - R_{B})^2$, which gives $R(E) = \frac{1}{1 + \exp{[\frac{2\pi}{\hbar \omega}( E- B)}]}$, where $\hbar\omega$ measures the curvature of the assumed inverted parabolic barrier, $\omega^2 = \frac{1}{m}\frac{d^{2}V_{inv}}{dr^2}|_{r = R_{B}}$, where $m$ is the mass of the atom. This procedure is quite common in the theory of nuclear fission \cite{HW53}. By fitting the data points for $R(E) \leq \frac{1}{2}$ we easily obtain the values of the height ( which can also be read off from the condition $R(E) = \frac{1}{2}$ ) and, more importantly, the curvature of the invisible barrier. The height and curvature parameter of the barrier for the data on $Ne$ atoms reflection off a silican wall, \cite{Sh01} are, respectively, $B = 1.03 nK$ and $\frac{d^2V(r)}{dr^2}|_{r =R_{B}} = 0.011 \frac{nK}{10^{-14}m^{2}}$, and on a BK7 glass wall, \cite{Sh01}, $B = 0.05 nK$ and $\frac{d^2V(r)}{dr^2}|_{r =R_{B}} = 0.015 \frac{nK}{10^{-14}m^{2}}$. In the case of the $Na$ BEC reflection data of \cite{Pas06} we find for these parameters the values, $B = 23.7 nK$ and $\frac{d^2V(r)}{dr^2}|_{r =R_{B}}= 17.0 \frac{nK}{10^{-14}m^{2}}$. Clearly the barrier for BEC reflection is more than an order of magnitude higher and thinner than for single atom reflection.\\

\begin{figure}
\centering
\includegraphics*[width=0.65\columnwidth]{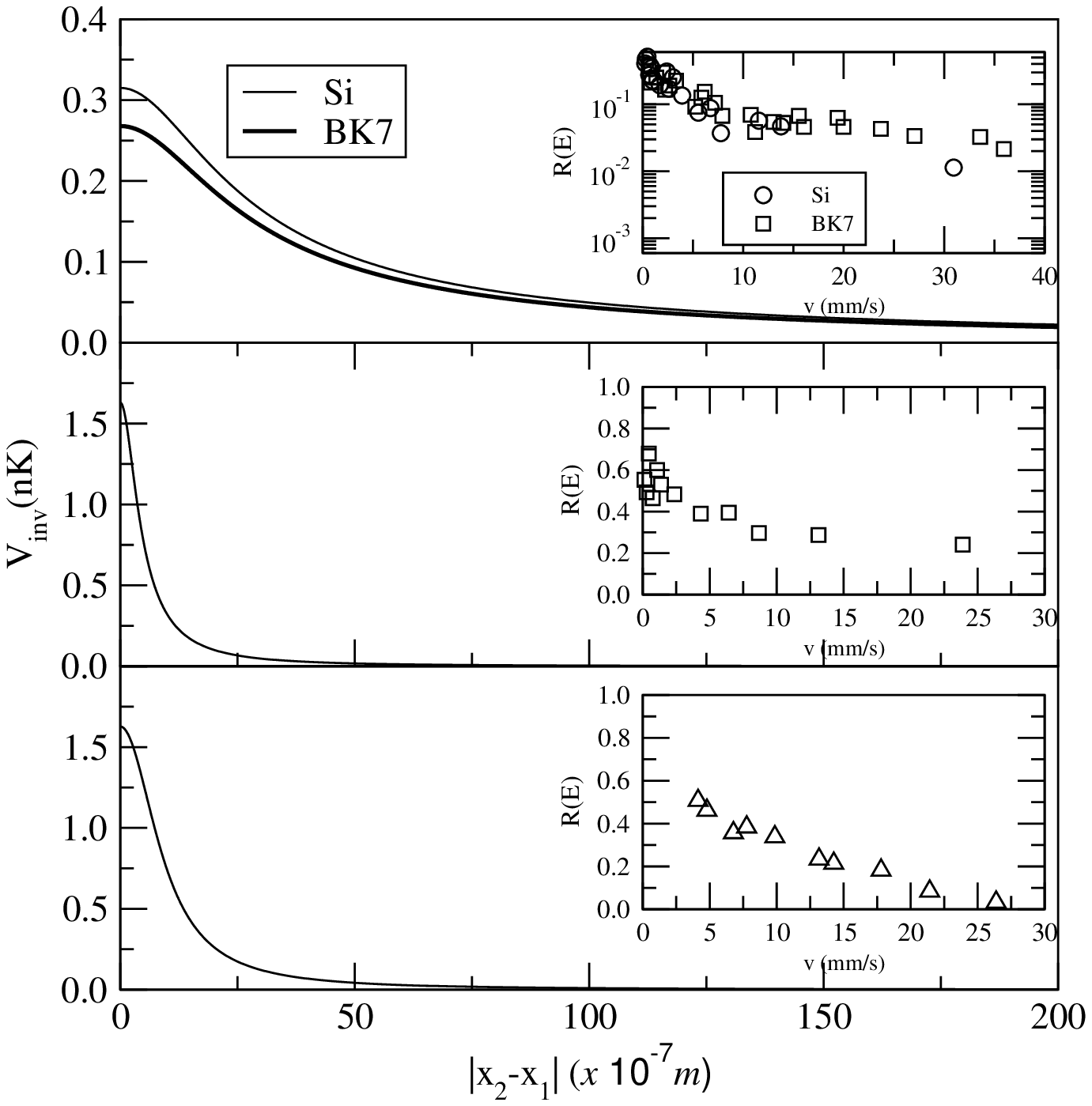}
\caption{The invisible barrier vs. its thickness for the data of \cite{Sh01, Pas06}. The experimental reflectivities were adjusted with a simple function and used in Eq. (6). Insets: Fig 1a.,
The reflectivity data versus the normal incident velocity on
the Si(1,0,0) surface \cite{Sh01}. Fig 1b.,
The reflectivity data versus the normal incident velocity on
the BK7 glass surface, \cite{Sh01}. Fig 1c.,
Reflectivity of sodium BEC data versus incident velocity. Data points, \cite{Pas06},
correspond to a sodium BEC confined in a magnetic trap with trap frequencies
$2\pi(2.0,2.5,8.2)$ Hz (full squares) and $2\pi(4.2,5.0,8.2)$ Hz
(open triangles). See text for details.}
\label{jos11}
\end{figure}

The Abel formula was employed using the available data for $R(E) > \frac{1}{2}$ and extrapolating these to lower energies using an appropriate fitting function. The fit formula which we have employed is of the general form $R(E) =\exp{[-C E^{D}]}$. Such a functional form reproduces the over all trend of the data of \cite{Sh01} and to a large extent those of the BEC reflectivity of \cite{Pas06}. In the case of reflection of neon atoms from a silicon wall of \cite{Sh01}, we find $C= 0.81$ and $D=0.71$, whereas for reflection of neon atoms from BK7 glass wall, $C= 0.92$, $D=0.29$. In the case of the sodium BEC reflection from a silicon wall \cite{Pas06}, we obtained $C=0.23$ and $D=0.39$. This latter case could account for the small reflectivity (higher velocities) and misses altogether the observerd saturation at smaller velocities. The inversion alluded to above would only supply the barrier value as a function of its thickness, $V_{inv} ( x_{1}-x_{2})$. Such potentials are shown in Fig. \ref{jos11}. The barrier heights and curvatures obtained from the inversion procedure are: $B = 1.03 nK$ and $\frac{d^2V(r)}{dr^2}|_{r =R_{B}} = 0.0086\frac{nK}{10^{-14}m^{2}}$ for neon on silicon wall,\cite{Sh01}, $B = 0.048 nK$ and $\frac{d^2V(r)}{dr^2}|_{r =R_{B}} = 0.0108 \frac{nK}{10^{-14}m^{2}}$ for neon on BK7 glass wall, \cite{Sh01}, and $B = 23.71 nK$, $\frac{d^2V(r)}{dr^2}|_{r =R_{B}} = 17.64 \frac{nK}{10^{-14}m^{2}}$ for sodium BEC reflection from silicon wall \cite{Pas06}. The values of the curvatures are quite consistent with those obtained using the parabolic approximation above.\\

\begin{figure}
\centering
\includegraphics*[width=0.65\columnwidth]{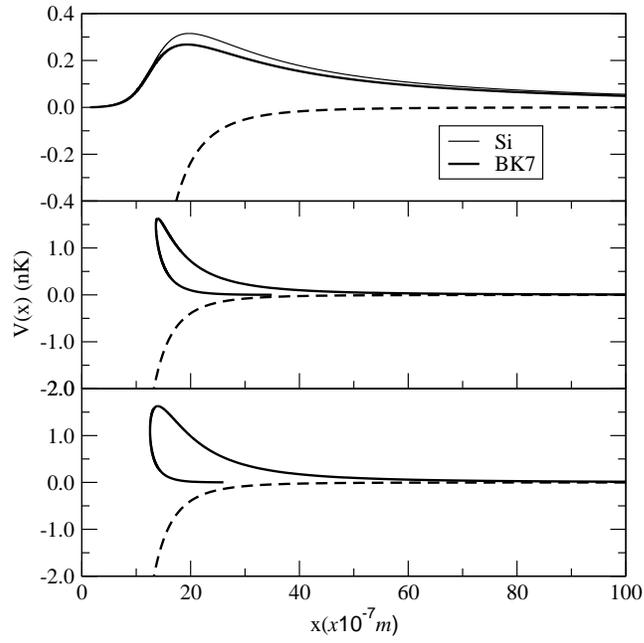}
\caption{ The invisible barrier vs. the distance, $x$, for $Ne$ reflection from a silicon wall ( at $x =0.0$ ), from a BK7 glass wall,  \cite{Sh01} and for $Na$ BEC reflection from a silicon wall \cite{Pas04, Pas06}. The dashed curve indicates the Casimir-Polder potential. See text for details.}
\label{jos21}
\end{figure}

The result of the calculation of the BEC $V_{inv}$ using the time-dependent Gross-Pitaevskii equation is given below. Before we discuss these results, it would be useful to obtain the invisible barrier as a function of the distance between the reflected entity and the silicon wall. For this purpose we need another equation involving the two turning points, $x_{1}$ and $x_{2}$. 
One possible suggestion is to relate their product
to the square of the virtual turning point, $r(E)$, related to the Casimir-Polder potential $V_{CP}$, namely $E = 
V_{CP}(r(E))$, and $r(E)^{2} = x_{1}\cdot x_{2}$. Through this relation we get the real turning points $x_{1}$ and 
$x_{2}$
separately, and accordingly we can construct the invisible quantum barrier as a function of the distance from the wall. This procedure is not unique as a different relation involving the turning points and the Casimir-Polder potential would yield a different shape for the invisible barrier. Not having available 
another plausible relation, we proceed and use the product relation above. 
The result of such a calculation is presented in Fig \ref{jos21}. We see clearly that the barrier is a double valued 
function in both axes. This feature may be a reflection of the use of an energy-dependent relation between the turning points and $r(E)$. It would be certainly of value to explore other relations, such as $x_{1} + x_{2} = 2\cdot r(E)$.\\

\bigskip

\section{Invisible Quantum Barrier for BEC}
We now turn to the study of quantum reflection of an one dimensional Bose-Einstein condensate from the silicon surface. Several theoretical studies of 1D BEC have been published ( see, e.g. \cite{Druten, Bouch}). It has been argued that by lowering the dimension to 1D, the transition temperature for BEC formation is increased. These systems are subtle and exhibit features not encountered in 2D or 3D. Bearing in mind the intrinsic differences between the 1D and 3D BEC's, we shall, nevertheless, discuss the quantum reflection of 1D BEC for the purpose of simplicity.  Would the condensate suffer reflection just as a single atom? In a way, to answer this question is in line with a broader one connected with the quantum behaviour of the motion of mesoscopic systems in general. One is reminded here with the pioneering work of M. Arndt \cite{Arn99, Arn05} on interference pattern of $C_{60}$ and other heavy molecules as they pass through a grating. As in \cite{Pas06}, we consider N $^{23}Na$ atoms Bose-Einstein condensed into the ground state in an anisotropic trap, $\Omega_{\rho}\gg \Omega_x$, where $\Omega_{\rho}$ is the trap frequency in the transverse directions and $\Omega_x$ the frequency in x-direction.  Both the quantum reflection and the interaction between atoms will not  excite the motion in the transverse direction. With these assumptions, the condensate is described by the one-dimensional Gross-Pitaevskii (GP) equation \cite{PMH98}. In fact one dimensional BEC has been formed and studied in \cite{Gor01, Weiss, Druten1, Aspect}. The 1D GP equation
we shall solve here is \cite{PMH98}

\begin{equation}
 i\hbar\frac{\partial\psi}{\partial t}=\left[-\frac{\hbar^2}{2M}\frac{d^2x}{dx^2}+V({\bf x}) +g|\psi|^2\right]\psi
\end{equation}
where $g=N\times g_{1D}$ with N the number of atoms and $g_{1D}=2\hbar\Omega_{\rho}a$, where
$a$ is the $s$-wave scattering length, which can be tuned through the Feshbach resonance \cite{Tim99}. The condition of validity of the above formula for $g_{1D}$ is: $a \ll \sqrt{\frac{\hbar}{m\Omega_{\rho}}}$ \cite{Olshanii}.
Notice that the number of atoms appears in the strength of the interaction $g$ as the wave function has been normalized to unity,

\begin{equation}
\int |\psi|^2dx=1.
\end{equation}



The external potentials include the harmonic trap in x-direction and  Casimir-Polder potential induced by a silicon 
surface~\cite{Pas06, Pas04}, $V({\bf x})=V_{trap}+V_{cp}$. The harmonic trap is $V_{trap}=\hbar\Omega_x^2 x^2/2$. In 
this paper, the trap frequency is set to be $\Omega_x=2\pi\times3.5 Hz$, in the range of values of recent 
experiments~~\cite{Gor01, Pas06, Pas04}. The Casimir-Polder potential has the form of Eq. (4)
. Similar to ~\cite{Sco05}, the model 
Casimir-Polder potential is separated into two regions.  If $x<d-\delta$, we have the normal Casimir-Polder potential. 
Here $\delta=0.15\mu\text{m}$ is a small offset to silicon surface.  When   $x>d-\delta$, we use a complex potential, 
$V_{c}=V_{cp}(\delta)-i(x-\delta)V_i$ modeling adsorption at the surface with $V_i=1.6\times10^{-26}\text{J$\cdot$ m}^{-1}$.  



\begin{figure}
\centering
\includegraphics*[width=0.65\columnwidth]{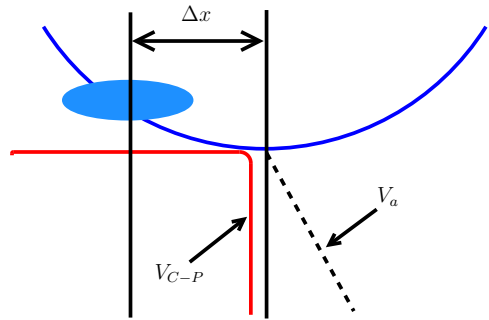}
\caption{(Color online.) A schematic diagram of the potentials.  The blue curve is the harmonic trap. A ground state condensate was created in the trap. The red curve is the Casimir-Polder poential. In the numerical calculation, we set a finite offset to the surface(see text). The black curve indicates the imagnary part of the absorbing/adsorbing potential. If atoms reach this region, they will be adsorbed at the surface or inelastically scattered away from the trap.  
\label{dia1}
}
\end{figure}

Recently experiments on the reflection of condensates were realized in three dimensional traps with number of 
condensed 
atoms reaching $~10^6$~\cite{Pas06, Pas04}. However, the number of atoms which can be condensed in one dimensional 
trap is 
reduced due to thermal and quantum fluctuations.  We use the number of atoms in the BEC of a recent experiment~\cite{Gor01}, $N_{1d}=1.5\times10^4$. At first we "prepare" a ground state of condensate in the harmonic trap.  
The center of trap was shifted to the Si surface suddenly by a distance $\Delta x$. Therefore the condensate  is 
suddenly put  
at a high potential: $\Delta V=M\Omega_x^2 \Delta x^2/2$. 
This induces the condensate to travel towards the surface, see Fig. \ref{dia1}. 
At the surface, the incident velocity is approximated by $v_x\approx \Omega_x \Delta x$.  
We use a fourth-order Runge-Kutta method in the interaction picture to perform the time propagation of the solution of 
the Gross-Pitaevskii equation. After one complete reflection, 
we define reflection probability as the number of left-traveling atoms divided by the number 
of incident atoms.  Different from the 
single atom reflected by the surface, the 
condensate has a finite spatial extension. This yields a minimal but nonzero velocity of incident condensate. In 
addition, during the reflection process, atoms are lost continuously. The nonlinear term in the Gross-Pitaevskii 
equation is also reduced accordingly since it is  proportional to the total number of  atoms in the condensate. 
 
\begin{figure}
\centering
\includegraphics*[width=0.65\columnwidth]{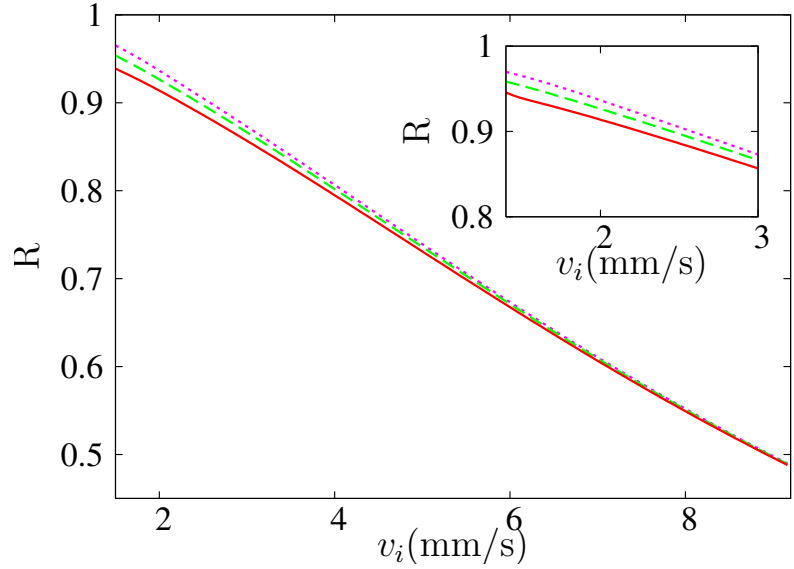}
\caption{(Color online.) Quantum reflection of one dimensional 
sodium Bose-Einstein condensate from a silicon surface. Velocity is measured in $mm/s$. 
Three values of the atom-atom interaction strength 
were considered: $g_s = 0.0 $  (dashed line), $g_s = 50.0$ 
(dashed dotted line) and $g_s = 200.0$ (solid line). See text for details.}
\label{r01}
\end{figure}

The simulation is presented in Fig. \ref{r01}. When the velocity is reduced, the reflection probability approaches 
unity. However, due to the facts mentioned above, the velocity can not reach zero. So, in principle, we get a less-than unity maximum reflection, similar to the experimental results of \cite{Pas06}. With the reflectivity calculated, we can now obtain the invisible barrier following the procedure detailed above.

\begin{figure}
\centering
\includegraphics*[width=0.95\columnwidth]{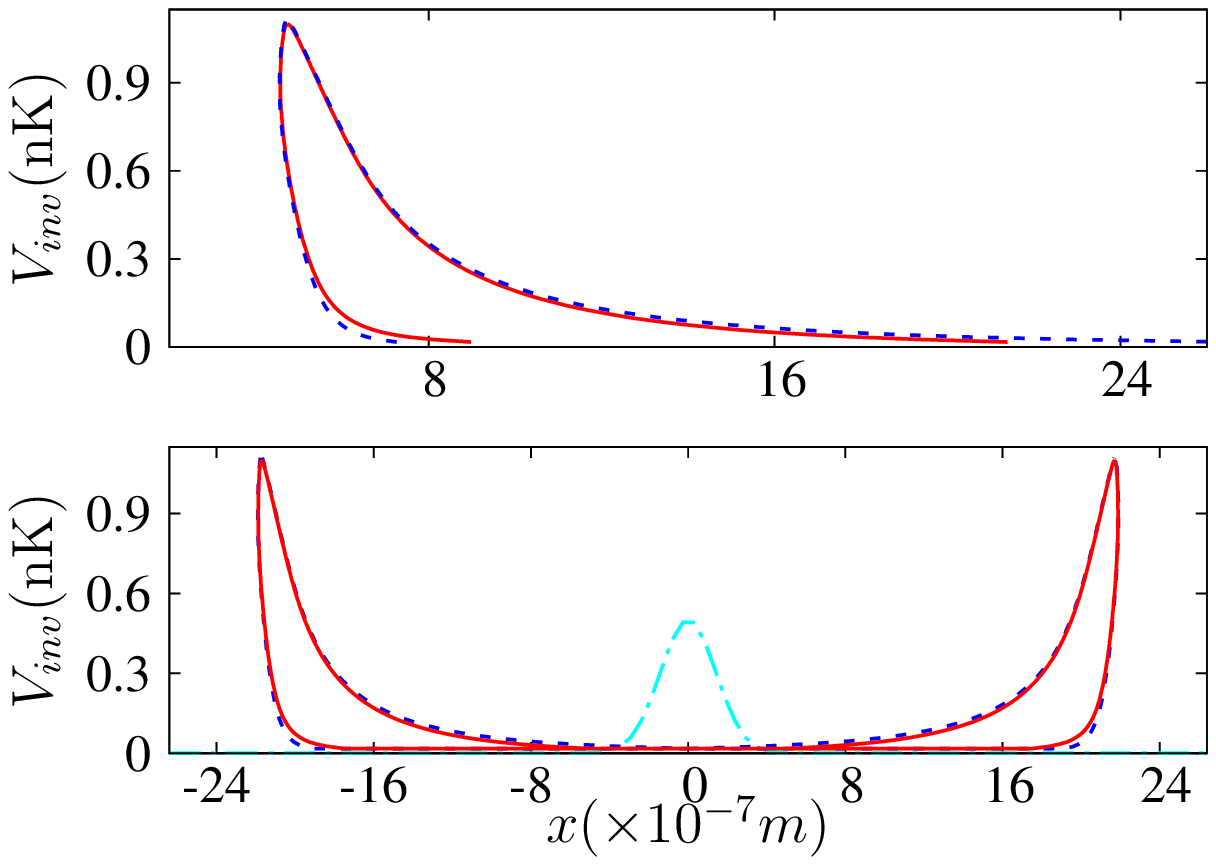}
\caption{(Color online.) Top: The invisible barrier of a silicon surface ( at $x =0.0$ ) for a sodium Bose 
condensate for two values of the  interaction strength, $g_s = 0.0$ (dashed) and $200.00$ (full). Here we use 
the units: distance in meter and energy in $nK$. Bottom: The same as the above figure but with  two silicon walls situated at $x= ( -26, +26 ) \times 10^{-7}m$, schematically showing the 
quantum reflection trap. The Gaussian form at the center is the BEC. See text for details.}
\label{invbecA}
\end{figure}

We use the Eq.(6) to get the barrier at different thicknesses. The corresponding $V_{inv}(r)$ is shown in 
Fig. \ref{invbecA} and it resembles very well the one obtained from the data of \cite{Pas06}, Fig. \ref{jos21}. 
This attests to the consistency of our calculation. More importantly, it shows clearly that a BEC behaves very much like a single atom,
albeit with a finite size. The size effect is felt when the BEC comes very close to the wall, as has already been mentioned.
We have also investigated the effect of the strength of the atom-atom interaction, $g$,
on the reflectivity of the BEC. In the numerical simulation, we scale the energy and length with ground state harmonic oscillator energy, $E_0=\hbar\Omega_x$ and harmonic length $l_0=\sqrt{\hbar/m\Omega_x}$. In this way, we solve the equation with scaled interaction strength $g_s = 0.0$ ( ideal gas ), $g_s=50$, and $g_s = 200.00$. The resulting reflectivity 
and the 
corresponding $V_{inv}(r)$ shows appreciable sensitivity to the value of $g$, and accordingly on the 
scattering length, $a$, Fig. \ref{invbecA}, which is comparable 
to the finite size effect alluded to above. We have also calculated the parameters of the barrier using the inverted
parabola approximation of Hill-Wheeler. We find, for the height $B= 111 nK$ and the curvature $\frac{d^2V(r)}{dr^2}|_{r =R_{B}} = 195(g = 0.0), 141 (g = 200) \frac{nK}{10^{-14}m^{2}}$. These values are to be compared to the ones extracted from the data, mentioned
above. The calculated ones are a factor 5-6 larger than the ones obtained from the data. This discrepancy could be traced
to the several limitations inherent in our 1D GP simulation.

\begin{figure}
\centering
\includegraphics*[width=0.65\columnwidth]{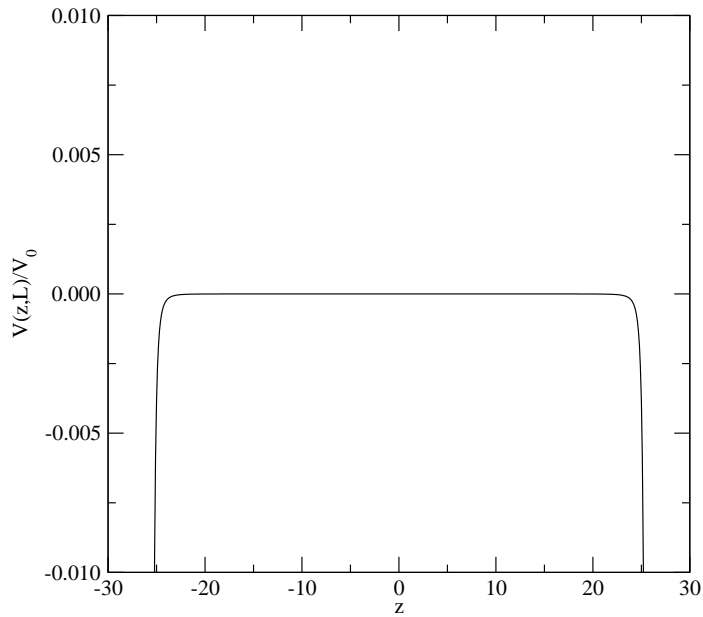}
\caption{The profile of the wall-atom-wall potential according to Eq. (14). The walls are situated at ( -26, +26)x $10^{-7}$ m}
\label{WAW}
\end{figure}

It has been recently suggested that that one could trap cold atoms and BEC by quantum reflection from two walls 
\cite{Jur05, Jur08}. Our findings above give a specific mechanism for such trapping. The cold atom or BEC will be in 
the confines of the two invisible barriers that such walls will generate. As long as the velocity of the atom or the 
BEC does not exceed the top of the invisible barrier, they will be trapped in free space without touching the walls in 
accordance with the conclusions of \cite{Jur05, Jur08}. The wall-atom-wall CP potential can be written as \cite{Kha97}

\begin{equation}
V_{WAW}(z, L) = \frac{1}{L^4} \Big [ \frac{1}{360} - \frac{ 3 - 2 \cos^2 (\frac{\pi z}{L})}{8 cos^4(\frac{\pi z}{L})} \Big]
\end{equation}
where $z$ is the distance of the atom from the mid-point between the walls, taken to be separated by $L$. The profile of the above WAW potential is shown in Fig. \ref{WAW}. For the purpose of completeness we calculate the corresponding double invisible barrier when the BEC is placed between two walls.

A schematic 
picture 
of the confining double-barrier in the case of the sodium BEC is shown in Fig. \ref{invbecA}. One sees clearly that the double-barrier defines the confining space to be the interval ( -22, +22) x $10^{-7}$ m, while the actual walls are situated
at (-26, +26) x $10^{-7}$ m. The space between the barrier and the wall is completely inaccessible. This, however, depends on the energy, as the invisible barrier is energy-dependent. In the case of sodium BEC \cite{Pas06}, the extracted critical velocity which the BEC must have in order to reach this forbidden region corresponds to an energy of about $24.0 nK$. Our
1D GP calculation gave a height which is 5 times greater. This is due to the inherent limitation of the 1D simulations. We should also warn the reader that our confining box-like potential was constructed with using the plausible condition on the turning points, $x_1\times x_2 = r(E)^2$, with $r(e)$ being the virtual turning point associated with the Casimir-Polder potential. Another condition would give a different confining region.

\bigskip

\section{Conclusions}
In conclusion we have inverted the recent data on reflectivity of neon atoms and of sodium Bose-Einstein condensates
to obtain the invisible quantum barrier. We have found that this new object is double valued in both axes. The height and curvature of the invisible barrier were determined from the data. The value of the height we obtain form the data of \cite{Pas06} is $B = 24 nK$, while for the neon atoms of \cite{Sh01} it is $1.0 nK$. The former is high enough that makes the trapping of BEC quite feasible at energies smaller than $B$. Such trapping could be potentially of great value in the fabrication of atom laser devises, as the BEC will maintain its integrity without suffering from absorption-adsorption at the walls. 
Further, we have verified that a BEC, though mesoscopic and contains millions of atoms, behaves, to a large extent, like a single, albeit large, atom when reflected from a solid wall. This finding corroborates  those of \cite{Pas06,Pas04} on the persistence of quantum behavior of large mesoscopic objects and is in line with the results of the recent grating diffraction experiments on buckyballs and other large molecules \cite{Arn99,Arn05}.

\bigskip

We thank Drs. H. Friedrich and J. M. Rost for discussion. We also thank W. Ketterle's group at MIT for supplying the data points mentioned in \cite{Pas06}. This work is partly supported by the Brazilian agencies, CNPq and FAPESP. M. S. Hussein is the Martin Gutzwiller Fellow 2007/2008.

\newpage

\ \\

\ \\

\ \\

\newpage


\begin{thebibliography}{101}
\bibitem{Na83} V. U. Nayak, D. O. Edwards, and N. Masuhara, Phys. Rev. Lett. \textbf{50}, 990 ( 1983).
\bibitem{Be89} J. J. Berkhout, O. J. Luiten, I. D. Setija, T. W. Hijjmans, T. Mizusaki, and J. T. M. Walraven, 
Phys. Rev. Lett. \textbf{63}, 1689 (1989).
\bibitem{Yu93} I. A. Yu, J. M. Doyle, C. L. Cesar, D. Kleppner and T. J. Greytak, Phys. Rev. Lett. \textbf{71}, 1589 (1993).
\bibitem{Sh01} F. Shimizu, Phys. Rev. Lett. \textbf{86}, 987 (2001).
\bibitem{Pas04} T. A. Pasquini, Y. Shin, C. Sanner, M. Saba, A. Schirotzek, D. E. Pritchard and W. Ketterle, Phys. 
Rev. Lett. \textbf{93}, 223201 (2004)
\bibitem{Pas06} T. A. Pasquini, M. Saba, G. -B. Jo, Y. Shin, W. Ketterle, D. E. Pritchard, T. A. Savas and N. Mulders, 
Phys. Rev. Lett. \textbf{97}, 093201 (2006).
\bibitem{CP48} H. Casimir and D. Polder, Phys. Rev. \textbf{73}, 360 (1948).
\bibitem{Sco05} R. G. Scott, A. M. Martin, T. M. Fromhold and F. W. Sheard, Phys. Rev. Lett. \textbf{95}, 073201 (2005).
\bibitem{Lin04} Yu-Ju Lin, I. Teper, C. Chin, and V. Vulti\'c, Phys. Rev. Lett. \textbf{92}, 050404 (2004).
\bibitem{MH96} N. T. Maitra and E. J. Heller, Phys. Rev. \textbf{A 54}, 4763 (1996).
\bibitem{Hen96} C. Henkel, C. I. Westbrook, and A. Aspect, J. Opt. Soc. Am. \textbf{B 13}, 233 (1996).
\bibitem{Fri02} H. Friedrich, G. Jacoby, and C. G. Meister, Phys. Rev. \textbf{A 65}, 032902 (2002)
\bibitem{Seg97} B. Segev, R. C\^ote' and M. G. Raizen, Phys. Rev. \textbf{A 56}, R3350 (1997).
\bibitem{Mar97} M. Marinescu, A. Dalgarno and J. F. Babb, Phys. Rev. \textbf{A 55}, 1530 (1997).
\bibitem{Kha97} P. Kharchenko, J. F. Babb and A. Dalgarno, Phys. Rev. \textbf{A 55}, 3530 (1997).
\bibitem{Kem35} E. C. Kemble, Phys. Rev. \textbf{48}, 549 (1935).
\bibitem{LL69} L. D. Landau and E. M. Lifschitz, $Mechanics$(Pergamon, New York, 1969)
\bibitem{CG78}  M. W. Cole and R. H. Good, Jr., Phys. Rev. \textbf{A 18}, 1085 (1978).
\bibitem{HW53} D. L. Hill and J. A. Wheeler, Phys. Rev. \textbf{89}, 1102 (1953).
\bibitem{BKN83} A. B. Balantekin, S. E. Koonin and J. W. Negele, Phys. Rev. \textbf{C 28}, 1565 (1983).
\bibitem{Lang37} R. E. Langer, Phys. Rev. {\bf 51}, 669.
\bibitem{Druten} W. Ketterle and N. J. van Druten, Phys. Rev. {\textbf A 54}, 656 (1996).
\bibitem{Bouch} I. Bouchoule, K. V. Kheruntsyan, and G. V. Shlyapnikov, Phys. Rev. {\textbf A 75}, 031606 (2007).
\bibitem{Arn99} M. Arndt et al., Nature \textbf{401}, 680 (1999).
\bibitem{Arn05} M. Arndt, K. Homberger, and A. Zeilinger, Physics World (March), 35-40 (2005).
\bibitem{PMH98} V. M. P\'erez-Garc\'ia, H. Michinel and H. Herrero, Phys. Rev. \textbf{A 57}, 3837 (1998).
\bibitem{Tim99} E. Timmermans, P. Tommasini, M. Hussein and A. Kerman, Phys. Rep. \textbf{315}, 199 (1999).
\bibitem{Olshanii} M. Olshanii, Phys. Rev. Lett., \textbf{81}, 938 (1998).
\bibitem{Gor01} A. G\"orlitz et al., Phys. Rev. Lett. \textbf{87}, 130402 (2001).
\bibitem{Weiss} T. Kinoshita, T. Wenger and D. S. Weiss, Phys. Rev. Lett. {\bf 95}, 190406 (2005)
\bibitem{Druten1} A. H. van Amerongen et al., Phys. Rev. Lett. {\bf 100}, 090402 (2008)
\bibitem{Aspect} D. Cl\'ement et al., Phys. Rev. Lett. {\bf 95}, 17409(2005)
\bibitem{Jur05} A. Jurisch and H. Friedrich, Phys. Lett. \textbf{A 349}, 230 (2005).
\bibitem{Jur08} A. Jurisch and J. M. Rost, ArXiv:0801.2514.
\end{thebibliography}
\end{document}